# A Class of Multipartite Entangled States Based on State Transitions

Jehn-Ruey Jiang

Department of Computer Science and Information Engineering
National Central University, Taoyuan City, Taiwan
jrjiang@g.ncu.edu.tw

June 4, 2026

**Abstract**

We introduce Transition states (T states), denoted by $|T_k^n\rangle$, as a class of multipartite entangled states characterized by a fixed number of state transitions between adjacent qubits. These states form equal-amplitude superpositions over all states with a specified transition count. Unlike Bell states based on two-qubit correlations, GHZ states characterized by global correlations among all qubits, and W and Dicke states based on fixed numbers of qubit excitations, T states are defined by transition counts along an ordered sequence of qubits. We prove that T states are unitarily equivalent to Dicke states through a chain of CX (controlled-X) operations, thereby establishing a direct correspondence between transition-based and excitation-based representations of multipartite entanglement.

# 1 Introduction

The term "entanglement" was coined by Schrödinger in 1935 [1] in response to the EPR paradox [2]. Schrödinger referred to entanglement as "the characteristic trait of quantum mechanics," whereas Einstein described it as "spooky action at a distance." Entanglement arises when multiple quantum systems (or particles) become correlated such that the state of each system cannot be described independently of the others, regardless of their spatial separation. When one part of an entangled state is measured, the states of the remaining parts are determined accordingly. This non-classical correlation is widely recognized as a key resource in quantum information science and is believed to play an important role in enabling computational advantages of quantum systems over their classical counterparts [3].

In quantum information processing, physical quantum systems are typically abstracted as qubits, which serve as the fundamental units of quantum information. Under this abstraction, entangled quantum systems can be represented as entangled qubit states, enabling their manipulation and implementation on quantum computing platforms. Such entangled states have been extensively applied in a variety of domains, including quantum cryptography [4], quantum dense coding [5], quantum teleportation [6], quantum secret sharing [7], entanglement swapping [8], quantum networks [9], quantum internet [10], distributed quantum computing [11, 12], quantum metrology and sensing [13–16], and quantum error correction [17–19].

Several canonical classes of multipartite entangled states have been studied in the literature. The Bell states [20] represent the simplest form of bipartite entanglement. For multi-qubit systems, the Greenberger–Horne–Zeilinger (GHZ) states [21] describe multipartite entangled states in which all qubits are globally correlated, forming a superposition of the all-$|0\rangle$ state and the all-$|1\rangle$ state. The W states [22] describe multipartite entangled states in which exactly one qubit is in $|1\rangle$ and the remaining qubits are in $|0\rangle$, forming a superposition of all such single-excitation basis states. The Dicke states [23] generalize the W states by describing multipartite entangled states in which a fixed number of qubits are in $|1\rangle$ and the remaining qubits are in $|0\rangle$, forming a superposition of all such basis states with the same number of excitations, i.e., states with a fixed Hamming weight.

Many combinatorial structures and physical systems are naturally characterized by transition patterns rather than excitation counts. Examples include sequences with a fixed number of state changes, boundary configurations, and optimization problems with transition-count constraints. This observation motivates the study of quantum states characterized by transition counts, providing a transition-based representation of multipartite entanglement that complements the excitation-based representation of entangled states.

In this paper, we propose a class of multipartite entangled states, called Transition states (or T states), denoted by $|T_k^n\rangle$. T states are characterized by a fixed number of state transitions between adjacent qubits and form equal-amplitude superpositions over all states with a specified transition count. They differ from the canonical classes of multipartite entangled states, which are typically constrained by fixed numbers of qubit excitations. In contrast, T states introduce a transition-based representation that captures the number of state changes along an ordered sequence of qubits, thereby providing an alternative and complementary perspective on structured quantum states.

We further establish that T states are in one-to-one correspondence with Dicke states, thereby revealing an intrinsic structural connection between transition-based and excitation-based representations. This correspondence not only provides new insights into the structure of multipartite entanglement but also enables the systematic construction of T states via transformations from Dicke states.

The remainder of this paper is organized as follows. Section 2 reviews related entangled states. Section 3 introduces the definition, examples, properties, and a theorem of T states. Section 4 concludes the paper and describes future research directions.

# 2 Related Work

## 2.1 Bell States

There are four Bell states, each of which is an entangled two-qubit state and represents the simplest and most fundamental form of quantum entanglement. The four Bell states are defined as

$$|\Phi^+\rangle = \frac{1}{\sqrt{2}}(|00\rangle + |11\rangle),$$

$$|\Phi^-\rangle = \frac{1}{\sqrt{2}}(|00\rangle - |11\rangle),$$

$$|\Psi^+\rangle = \frac{1}{\sqrt{2}}(|01\rangle + |10\rangle),$$

$$|\Psi^-\rangle = \frac{1}{\sqrt{2}}(|01\rangle - |10\rangle).$$

These states form an orthonormal basis for the two-qubit Hilbert space, commonly referred to as the Bell basis. Although Bell did not explicitly introduce the concept of the Bell basis, his 1964 analysis [20], which established what is now known as Bell's theorem, is based on the singlet state, namely the Bell state $|\Psi^-\rangle$.

Each of these states exhibits maximal entanglement, meaning that the measurement outcomes of one qubit are perfectly correlated (or anti-correlated) with those of the other qubit. As a result, the joint state cannot be written as a tensor product of individual qubit states, reflecting the intrinsically quantum nature of entanglement [17].

Bell states serve as essential resources in many quantum information processing applications, such as certain entanglement-based quantum cryptography schemes [4], quantum dense coding [5], and quantum teleportation [6]. They are also widely used in experimental demonstrations of quantum nonlocality via Bell inequality violations [24]. Their simple structure and perfect correlations or anti-correlations (in appropriate measurement bases) make them a cornerstone of both theoretical and experimental quantum information science.

## 2.2 GHZ States

The Greenberger–Horne–Zeilinger (GHZ) states [21] represent a fundamental class of multipartite entangled states that extend bipartite entanglement to multi-qubit systems. These states were originally introduced to explore non-classical correlations in quantum mechanics and to demonstrate a stronger form of quantum nonlocality beyond Bell-type inequalities.

For an $n$-qubit system, a GHZ state is defined as a superposition of two globally correlated basis states, namely the all-$|0\rangle$ state and the all-$|1\rangle$ state, given by

$$|\mathrm{GHZ}_n\rangle = \frac{1}{\sqrt{2}} \left( |0\rangle^{\otimes n} + |1\rangle^{\otimes n} \right).$$

For example, the 3-qubit GHZ state $|GHZ_3\rangle$ is given by $\frac{1}{\sqrt{2}} (|000\rangle + |111\rangle)$.

In GHZ states, all qubits are perfectly correlated in the sense that measuring any single qubit in the computational basis immediately determines the outcomes of all others in the same basis. This global correlation structure distinguishes GHZ states from other forms of multipartite entanglement and makes them a canonical example of highly entangled multipartite states [17].

GHZ states have been used to design protocols involving multi-party quantum communication and quantum secret sharing [7]. In addition, they have been shown to enable distributed control and non-local quantum operations across multiple distributed quantum computing nodes [11], and more recent works further identify them as a primitive for distributed fan-out operations that reduce circuit depth [12]. GHZ states have also been widely used in quantum metrology and sensing, enabling precision beyond classical limits. In particular, they support phase estimation and multiparameter metrology approaching the Heisenberg limit [13–15], and can enhance sensing performance under realistic noise conditions [16]. Quantum error correction encodes information into entangled states distributed across multiple qubits [17, 18]. In particular, GHZ states can be viewed as simple error-correcting codes related to repetition codes that protect against bit-flip errors. Recent experiments show that GHZ-like entangled structures enable error detection and fidelity enhancement [19].

## 2.3 W States

The W states [22] represent a fundamental class of multipartite entangled states that exhibit a distinct type of entanglement from GHZ states. Unlike GHZ states, which are characterized by global correlations, W states capture a form of distributed entanglement that remains robust under qubit loss or local measurements, in the sense that entanglement persists among the remaining qubits even after part of the system is removed or measured.

For an $n$-qubit system, a W state is defined as an equal superposition of all computational basis states in which exactly one qubit is in $|1\rangle$ and the remaining qubits are in $|0\rangle$, given by

$$|W_n\rangle = \frac{1}{\sqrt{n}} \sum_{j=1}^{n} |0\rangle^{\otimes(j-1)} |1\rangle |0\rangle^{\otimes(n-j)}.$$

For example, the 3-qubit W state $|W_3\rangle$ is given by $\frac{1}{\sqrt{3}} (|001\rangle + |010\rangle + |100\rangle)$.

In W states, the excitation is delocalized across all qubits, and the loss of one qubit does not completely destroy the entanglement among the remaining qubits. Thus, W states have been exploited as robust multipartite entanglement resources to construct noise-resilient quantum networks, distributed quantum systems, and quantum-information protocols. Examples include semi-quantum secret sharing protocols [25], in which a dealer distributes secret information among multiple agents who must cooperate to recover it. In semi-quantum secret sharing, only the dealer has full quantum capabilities, while the other participants perform only classical operations, reducing implementation complexity while preserving quantum security. Beyond secret sharing, W states have also been employed for anonymous transmission in noisy quantum networks, where they can tolerate one non-responsive node [26].

## 2.4 Dicke States

The Dicke states [23] represent a fundamental class of multipartite entangled states that generalize W states to configurations with multiple excitations. Unlike GHZ states, which are characterized by global correlations, and W states, which correspond to a single excitation distributed across qubits, Dicke states describe entangled structures in which a fixed number of qubits are in the $|1\rangle$ state, forming a broader family of symmetric entangled states.

For an $n$-qubit system with $k$ excitations, a Dicke state is defined as an equal superposition of all computational basis states with exactly $k$ qubits in $|1\rangle$ and $n-k$ qubits in $|0\rangle$, given by

$$|D_k^n\rangle = \frac{1}{\sqrt{\binom{n}{k}}} \sum_{\substack{x\in\{0,1\}^n \\ w(x)=k}} |x\rangle ,$$

where $w(x)$ denotes the Hamming weight of the $x$. This definition highlights that Dicke states are characterized by a global constraint on a fixed Hamming weight, i.e., a specified number of excitations. Note that a Dicke state $|D_k^n\rangle$ contains $\binom{n}{k}$ basis states. For example, the 3-qubit Dicke state $|D_2^3\rangle$ is given by $\frac{1}{\sqrt{\binom{3}{2}}}(|011\rangle + |101\rangle + |110\rangle)=\frac{1}{\sqrt{3}}(|011\rangle + |101\rangle + |110\rangle)$.

Since Dicke states can be regarded as a generalization of W states, they inherit the robustness properties of W states to some extent. In general, the degree of robustness of a Dicke state $|D_k^n\rangle$ depends on the excitation number $k$. On the one hand, states with $k=1$ or $k=n-1$, which correspond to W-type states, are robust against qubit loss due to their distributed entanglement structure [27]. On the other hand, balanced Dicke states with $k \approx n/2$ are often robust against noise and decoherence, as their symmetric structure and large number of excitations lead to enhanced collective interference and averaging effects that mitigate the impact of local perturbations. This property makes them particularly advantageous in quantum metrology and sensing applications [13].

In addition, Dicke states provide a natural framework for encoding combinatorial constraints, as their structure directly corresponds to fixing the Hamming weight of states. This property has been exploited in studies [28–31] on solving combinatorial optimization problems (COPs) with fixed-cardinality constraints. These studies initialized working qubits in the Dicke state and utilized Grover's quantum search algorithm [32] or related variants to solve NP-complete COPs with fixed-cardinality constraints, including the $k$-vertex cover problem and $k$-hitting set problem. With the Dicke state initialization, the effective search space of the $n$ working qubits can be reduced from $2^n$ to $\binom{n}{k}$. Since $\binom{n}{k} = \mathrm{O}(n^k)$ for the condition of $\min(k, n-k) \ll n/2$, the search space is reduced from exponential in $n$ to polynomial in $n$ when $k$ is fixed [28]. Moreover, as shown in [30], using Dicke state not only simplifies the oracle design of Grover's algorithm but also reduces the number of required Grover iterators. This potentially reduces the number of required qubits and quantum gates, as well as the overall circuit depth, which is particularly beneficial in the NISQ era [33].

# 3 Transition States

## 3.1 Definition

We define the transition state (or T state) $|T_k^n\rangle$ as an $n$-qubit quantum state that is the equal-amplitude superposition over all computational basis states $|y\rangle = |y_{n-1}y_{n-2}\cdots y_0\rangle$ satisfying

$$y_0 + \sum_{j=1}^{n-1}(y_{j-1} \oplus y_j) = k, \tag{1}$$

where $n$ and $k$ are integers satisfying $0 < n$ and $0 \le k \le n$, $y_i \in \{0,1\}, 0 \le i \le n-1$, and $\oplus$ denotes the XOR operation.

In this definition, the term $y_0$ accounts for the transition from an implicit reference state $|0\rangle$ to the current state of the first qubit, ensuring a consistent counting manner across all positions. The summation counts the number of state transitions between adjacent qubits.

In general, $|T_k^n\rangle$ is an equal-amplitude superposition of all $n$-qubit states with exactly $k$ transitions, where a transition is counted either when two adjacent qubits differ or when the first qubit is in the state $|1\rangle$ (i.e., $y_0 = 1$). There are $n$ possible transition positions, and choosing $k$ of them yields exactly $\binom{n}{k}$ valid states. Therefore, each basis state in $|T_k^n\rangle$ has amplitude $1/\sqrt{\binom{n}{k}}$.

As will be shown later, a T state $|T_k^n\rangle$ has a bijective mapping to a Dicke state $|D_k^n\rangle$. In particular, the mapping

$$x_0 = y_0, \quad x_j = y_{j-1} \oplus y_j, \quad j \ge 1 \tag{2}$$

establishes a bijection between a state $|y_{n-1}y_{n-2}\cdots y_0\rangle$ with a fixed transition count and a state $|x_{n-1}x_{n-2}\cdots x_0\rangle$ with a fixed Hamming weight.

As an illustrative example, consider the case of $|T_2^5\rangle$, which consists of all 5-qubit computational basis states satisfying

$$y_0 + \sum_{j=1}^{4}(y_{j-1} \oplus y_j) = 2.$$

Under this condition, there are exactly $\binom{5}{2} = 10$ valid basis states, namely

$$|00001\rangle,\ |00010\rangle,\ |00011\rangle,\ |00100\rangle,\ |00110\rangle,$$
$$|00111\rangle,\ |01000\rangle,\ |01100\rangle,\ |01110\rangle,\ |01111\rangle.$$

Thus, the T state $|T_2^5\rangle$ is given by

$$|T_2^5\rangle = \frac{1}{\sqrt{10}}\sum_{y\in\mathcal{S}_2^5}|y\rangle,$$

where $\mathcal{S}_2^5$ denotes the set of all valid states satisfying the constraint of two state transitions.

Similarly, we can obtain other T states for $n = 5$, as shown below.

$$|T_0^5\rangle = |00000\rangle.$$

$$|T_1^5\rangle = \frac{1}{\sqrt{5}}\big(|10000\rangle + |11000\rangle + |11100\rangle + |11110\rangle + |11111\rangle\big).$$

$$|T_3^5\rangle = \frac{1}{\sqrt{10}}\big( |10001\rangle + |10010\rangle + |10011\rangle + |10100\rangle + |10110\rangle + |10111\rangle + |11001\rangle + |11010\rangle + |11011\rangle + |11101\rangle \big).$$

$$|T_4^5\rangle = \frac{1}{\sqrt{5}}\big( |00101\rangle + |01001\rangle + |01010\rangle + |01011\rangle + |01101\rangle \big).$$

$$|T_5^5\rangle = |10101\rangle .$$

## 3.2 Theorem

The definition in Eq. (1) implicitly defines a bijection between T states and Dicke states. We have the following theorem.

**Theorem 1.** *Let* $U_{\text{chain}} = \text{CX}_{n-2,n-1}\text{CX}_{n-3,n-2}\cdots\text{CX}_{0,1}$. *Then,* $|T_k^n\rangle = U_{\text{chain}}|D_k^n\rangle$ *and* $|D_k^n\rangle = U_{\text{chain}}^\dagger|T_k^n\rangle$.

*Proof.* Recall that the Dicke state $|D_k^n\rangle$ is defined as the equal-amplitude superposition over all states $|x\rangle = |x_{n-1}x_{n-2}\cdots x_0\rangle$ of Hamming weight $k$. That is,

$$|D_k^n\rangle = \frac{1}{\sqrt{\binom{n}{k}}} \sum_{\substack{x\in\{0,1\}^n \\ w(x)=k}} |x_{n-1}x_{n-2}\cdots x_0\rangle ,$$

where $w(x)$ denotes the Hamming weight of $x$.

Consider the action of $U_{\text{chain}}$ on a computational basis state $|x_{n-1}x_{n-2}\cdots x_0\rangle$ to obtain the output state $|y_{n-1}y_{n-2}\cdots y_0\rangle$. Sequentially applying the CX gates of $U_{\text{chain}}$ ensures that the output state $|y_{n-1}y_{n-2}\cdots y_0\rangle$ satisfies

$$y_0 = x_0, \qquad y_j = x_0 \oplus x_1 \oplus \cdots \oplus x_j, \quad j \geq 1.$$

Equivalently,

$$x_0 = y_0, \qquad x_j = y_{j-1} \oplus y_j, \quad j \geq 1.$$

Therefore,

$$w(x) = x_0 + \sum_{j=1}^{n-1} x_j = y_0 + \sum_{j=1}^{n-1} (y_{j-1} \oplus y_j).$$

Hence, $w(x) = k$ if and only if the corresponding $y$ satisfies

$$y_0 + \sum_{j=1}^{n-1} (y_{j-1} \oplus y_j) = k.$$

Consequently, applying $U_{\text{chain}}$ to $|D_k^n\rangle$ yields

$$U_{\text{chain}}|D_k^n\rangle = \frac{1}{\sqrt{\binom{n}{k}}} \sum_{\substack{y\in\{0,1\}^n \\ y_0+\sum_{j=1}^{n-1}(y_{j-1}\oplus y_j)=k}} |y\rangle .$$

By definition, this state is exactly $|T_k^n\rangle$. Thus, we have

$$|T_k^n\rangle = U_{\text{chain}} |D_k^n\rangle .$$

Since $U_{\text{chain}}$ is unitary, it follows that

$$U_{\text{chain}}^\dagger |T_k^n\rangle = U_{\text{chain}}^\dagger U_{\text{chain}} |D_k^n\rangle = I |D_k^n\rangle = |D_k^n\rangle .$$

□

By Theorem 1, we establish a bijection between T states and Dicke states. However, unlike Dicke states, T states are not invariant under qubit permutations. The ordering of qubits is essential due to the sequential definition of transitions. As a result, permuting qubits generally leads to a different state. This contrasts with the permutation symmetry of Dicke states.

# 4 Conclusion

In this paper, we introduced T states as a class of multipartite entangled states characterized by a fixed number of state transitions between adjacent qubits. T states differ from canonical classes of entangled states, namely the Bell, GHZ, W, and Dicke states. Specifically, Bell states are based on two-qubit correlations, GHZ states are characterized by global correlations among all qubits, and W and Dicke states are based on fixed numbers of qubit excitations. In contrast, T states are based on fixed numbers of state transitions between adjacent qubits, providing an alternative and structured description of quantum superposition. We further established a bijective relationship between T states and Dicke states via a chain of CX operations.

T states encode structural information via transition counts, making them potentially useful in quantum metrology for estimating spatially varying parameters such as gradients or boundary locations. Their sensitivity to local changes could also suggest possible applications in quantum sensing, particularly for detecting edges, or non-uniform fields where transition patterns carry physical significance. In quantum error correction, the transition-counting property may enable the design of codes that detect or localize bit-flip errors by monitoring changes in the transition structure of the quantum state. However, these possible applications require further theoretical and experimental investigation.

In quantum search settings where constraints are naturally expressed through adjacency relationships, T states enable the search space to be directly restricted to feasible configurations, thereby potentially reducing the effective search space. Following the approaches in studies [28–31], we can prepare the initial state of the working qubits as T states and apply Grover's quantum search algorithm or its variants to tackle combinatorial optimization problems with fixed transition-count constraints. This can potentially reduce the required qubits, gates, and circuit depth, thereby improving the practicality of implementing Grover-based search on NISQ devices.

# References

[1] Erwin Schrödinger. Discussion of probability relations between separated systems. In *Mathematical proceedings of the cambridge philosophical society*, volume 31, pages 555–563. Cambridge University Press, 1935.

[2] Albert Einstein, Boris Podolsky, and Nathan Rosen. Can quantum-mechanical description of physical reality be considered complete? *Physical review*, 47(10):777, 1935.

[3] Richard Jozsa and Noah Linden. On the role of entanglement in quantum computational speed-up. *Proceedings of the Royal Society of London. Series A: Mathematical, Physical and Engineering Sciences*, 459(2036):2011–2032, 2003.

[4] Artur K Ekert. Quantum cryptography based on Bell's theorem. *Physical review letters*, 67(6):661, 1991.

[5] Charles H Bennett and Stephen J Wiesner. Communication via one-and two-particle operators on Einstein-Podolsky-Rosen states. *Physical review letters*, 69(20):2881, 1992.

[6] Charles H Bennett, Gilles Brassard, Claude Crépeau, Richard Jozsa, Asher Peres, and William K Wootters. Teleporting an unknown quantum state via dual classical and Einstein-Podolsky-Rosen channels. *Physical Review Letters*, 70(13):1895, 1993.

[7] Mark Hillery, Vladimír Bužek, and André Berthiaume. Quantum secret sharing. *Physical Review A*, 59(3):1829, 1999.

[8] Marek Żukowski, Anton Zeilinger, Michael A Horne, and Aarthur K Ekert. "event-ready-detectors" Bell experiment via entanglement swapping. *Physical review letters*, 71(26):4287, 1993.

[9] H Jeff Kimble. The quantum internet. *Nature*, 453(7198):1023–1030, 2008.

[10] Marcello Caleffi, Angela Sara Cacciapuoti, and Giuseppe Bianchi. Quantum internet: From communication to distributed computing! In *Proceedings of the 5th ACM international conference on nanoscale computing and communication*, pages 1–4, 2018.

[11] Anocha Yimsiriwattana and Samuel J Lomonaco Jr. Generalized GHZ states and distributed quantum computing. *arXiv preprint quant-ph/0402148*, 2004.

[12] Seng W Loke. On distributed quantum computing with distributed fan-out operations. *arXiv preprint arXiv:2601.14734*, 2026.

[13] Géza Tóth and Iagoba Apellaniz. Quantum metrology from a quantum information science perspective. *Journal of Physics A: Mathematical and Theoretical*, 47(42):424006, 2014.

[14] Stefano Pirandola, B Roy Bardhan, Tobias Gehring, Christian Weedbrook, and Seth Lloyd. Advances in photonic quantum sensing. *Nature Photonics*, 12(12):724–733, 2018.

[15] Zheshen Zhang, Chenglong You, Omar S Magaña-Loaiza, Robert Fickler, Roberto de J León-Montiel, Juan P Torres, Travis S Humble, Shuai Liu, Yi Xia, and Quntao Zhuang. Entanglement-based quantum information technology: a tutorial. *Advances in Optics and Photonics*, 16(1):60–162, 2024.

[16] Allen Zang, Tian-Xing Zheng, Peter C Maurer, Frederic T Chong, Martin Suchara, and Tian Zhong. Enhancing noisy quantum sensing by GHZ state partitioning. *arXiv preprint arXiv:2507.02829*, 2025.

[17] Michael A Nielsen and Isaac L Chuang. *Quantum Computation and Quantum Information*. Cambridge university press, 2010.

[18] Barbara M Terhal. Quantum error correction for quantum memories. *Reviews of Modern Physics*, 87(2):307–346, 2015.

[19] Google Quantum AI and Collaborators. Quantum error correction below the surface code threshold. *Nature*, 638:920–925, 2025. Published online: 9 December 2024.

[20] John S Bell. On the Einstein Podolsky Rosen paradox. *Physics Physique Fizika*, 1(3):195, 1964.

[21] Daniel M Greenberger, Michael A Horne, and Anton Zeilinger. Going beyond Bell's theorem. In *Bell's theorem, quantum theory and conceptions of the universe*, pages 69–72. Springer, 1989.

[22] Wolfgang Dür, Guifre Vidal, and J Ignacio Cirac. Three qubits can be entangled in two inequivalent ways. *Physical Review A*, 62(6):062314, 2000.

[23] Robert H. Dicke. Coherence in spontaneous radiation processes. *Physical Review*, 93(1):99, 1954.

[24] Ryszard Horodecki, Pawe l Horodecki, Micha l Horodecki, and Karol Horodecki. Quantum entanglement. *Reviews of modern physics*, 81(2):865–942, 2009.

[25] Chia-Wei Tsai, Chun-Wei Yang, and Narn-Yih Lee. Semi-quantum secret sharing protocol using W-state. *Modern Physics Letters A*, 34(27):1950213, 2019.

[26] Victoria Lipinska, Gláucia Murta, and Stephanie Wehner. Anonymous transmission in a noisy quantum network using the w state. *Physical Review A*, 98(5):052320, 2018.

[27] John K Stockton, John M Geremia, Andrew C Doherty, and Hideo Mabuchi. Characterizing the entanglement of symmetric many-particle spin-(1/2) systems. *Physical Review A*, 67(2):022112, 2003.

[28] Jehn-Ruey Jiang. Dicke state quantum search for solving the vertex cover problem. *Mathematics*, 13(18):3005, 2025.

[29] Jehn-Ruey Jiang and Qiao-Yi Lin. Improving Grover's algorithm with Dicke states. In *2025 IEEE International Conference on Consumer Electronics-Taiwan (ICCE-Taiwan)*, pages 581–582. IEEE, 2025.

[30] Jehn-Ruey Jiang and Qiao-Yi Lin. Dicke-state-initialized Grover's algorithm for solving the hitting set problem. *Engineering Proceedings*, 134(1):94, 2026.

[31] Jehn-Ruey Jiang. Solving the k-hitting set problem with Dicke state quantum search. *Engineering Proceedings*, 134(1):38, 2026.

[32] Lov K Grover. A fast quantum mechanical algorithm for database search. In *Proceedings of the twenty-eighth annual ACM symposium on Theory of computing*, pages 212–219, 1996.

[33] John Preskill. Quantum computing in the NISQ era and beyond. *Quantum*, 2:79, 2018.